\documentclass{article}
\pdfoutput=1

\usepackage[oldstyle]{hep-paper}

\usepackage{scalerel}							
\usepackage{subdepth}							

\usepackage{subfiles}
\usepackage{import}
\usepackage{appendix}							
\AtBeginEnvironment{subappendices}{%
  \addtocontents{toc}{\protect\addvspace{10pt}Appendices}
}

\usepackage{upgreek}
\usepackage{xcolor}
\usepackage{verbatim}

\usepackage{tabularx}							%
\usepackage{makecell}							%
\usepackage[export = true]{adjustbox}				%

\usepackage{dsfont}

\usepackage{tikz}
\usepackage{tikzscale}
\usepackage{tikz-feynman}

\usepackage{bibentry}

\allowdisplaybreaks[1] 							

\usepackage{tocloft}

\AtBeginEnvironment{tabular}{\tlstyle}

\acronym{QFT}{quantum field theory}[quantum field theories]
\acronym{QCD}{quantum chromodynamics}
\acronym{QED}{quantum electrodynamics}
\acronym{GD}{gluon dynamics}
\acronym{SM}{Standard Model}
\acronym{RGE}{renormalization group equation}
\acronym{VEV}{vacuum expectation value}
\acronym{CKM}{Cabibbo-\allowbreak Kobayashi-\allowbreak Maskawa}
\acronym{PMNS}{Pontecorvo–\allowbreak Maki–\allowbreak Nakagawa–\allowbreak Sakata}
\acronym[\chi PT]{cPT}{chiral perturbation theory}[chiral perturbation theories]
\acronym{NGB}{Nambu-\allowbreak Goldstone boson}
\acronym{PNGB}{pseudo Nambu-\allowbreak Goldstone boson}
\acronym{WZW}{Wess-\allowbreak Zumino-\allowbreak Witten}
\acronym{MC}{Maurer-\allowbreak Cartan}
\acronym{GM}{Gell-\allowbreak Mann}
\acronym{CS}{Chern-\allowbreak Simons}
\acronym{YM}{Yang-\allowbreak Mills}
\acronym{HNL}{heavy neutral lepton}
\acronym{ALP}{axion-like particle}
\acronym{LLP}{long-lived particle}
\acronym{EOM}{equation of motion}
\acronym{PI}{partial integration}
\acronym[1PI]{onePI}{one particle irreducible}
\acronym{BSM}{beyond the Standard Model}
\longacronym{WIMP}{weakly interacting massive particle}
\acronym{DM}{dark matter}
\acronym{UV}{ultraviolet}
\acronym{IR}{infrared}
\acronym{LEC}{low energy coefficient}
\acronym{LER}{low energy realisation}
\acronym{NP}{new physics}
\acronym{EM}{electromagnetic}
\acronym{EW}{electroweak}
\acronym{EWSB}{electroweak symmetry breaking}
\acronym[\overline{MS}]{MSb}{modified minimal subtraction}
\acronym{NDA}{naive dimensional analysis}
\acronym{DOF}{degree of freedom}[degrees of freedom]
\acronym{LO}{leading order}
\acronym{NLO}{next-to-leading order}
\acronym{NNLO}{next-to-next-to-leading order}
\acronym{LHC}{Large Hadron Collider}
\shortacronym{CMS}{compact muon solenoid}
\shortacronym{ATLAS}{a toroidal LHC apparatus}
\shortacronym{LHCb}{LHC beauty}
\shortacronym{NA}*{North Area experiment \textnumero~}
\shortacronym{KOTO}{K0 at Tokai}
\shortacronym{SHiP}{Search for Hidden Particles}
\shortacronym{MATHUSLA}{Massive Timing Hodoscope for Ultra-Stable neutraL pArticles}
\shortacronym{FASER}{ForwArd Search ExpeRiment}
\shortacronym[CODEX-b]{CODEXb}{compact detector for exotics at LHCb}
\shortacronym{CP}{charge-parity}
\acronym{PET}{\emph{portal effective theory}}[\emph{portal effective theories}]
\acronym{EFT}{effective field theory}[effective field theories]
\acronym{SMEFT}{Standard Model effective field theory}[Standard Model effective field theories]
\acronym{HEFT}{Higgs effective field theory}[Higgs effective field theories]
\acronym{LEFT}{light effective field theory}[light effective field theories]
\acronym{SCET}{soft-collinear effective theory}[soft-collinear effective theories]
\acronym{HQET}{heavy quark effective theory}[heavy quark effective theories]
\acronym{NRQCD}{non-relativistic quantum chromodynamics}
\acronym{NRQED}{non-relativistic quantum electrodynamics}
\acronym[2HDM]{THDM}{two Higgs-doublet model}
\acronym{SUSY}{supersymmetry}
\acronym{SUGRA}{supergravity}
\acronym{MSSM}{minimal supersymemtric Standard Model}
\acronym{NMSSM}{next-to-minimal supersymemtric Standard Model}
\longacronym{CC}{charged current}
\longacronym{NC}{neutral current}
\longacronym{LE}{low energy}[low energies]
\longacronym{HE}{high energy}[high energies]
\acronym{BR}{branching ratio}

\acronyms{Omit}

\acronym{SNF}{Schweizerischer Nationalfonds zur Förderung der wissenschaftlichen Forschung}
\acronym[exp]{ex}{experimental}
\acronym{lat}{lattice}
\acronym{UCLouvain}{Université catholique de Louvain}
\acronym{FSR}{Fonds Spéciaux de Recherche}

\crefname{type}{type}{types}
\crefalias{inlinelisti}{type}
\crefalias{enumi}{type}

\crefformat{footnote}{#2\footnotemark[#1]#3}

\eqcrefname{lag}{Lagrangian}
\eqcrefname{set}{stress-energy tensor}
\eqcrefname{cur}{current}
\eqcrefname{sym}{symmetry}
\eqcrefname{bilinear}{quark bilinear}

\newcommand{\makediff}[2]{\expandafter\NewDocumentCommand\csname#1\endcsname{ogd()}{\IfNoValueTF{##2}{\IfNoValueTF{##3}{#2\IfNoValueTF{##1}{}{^{##1}}}{\mathinner{#2\IfNoValueTF{##1}{}{^{##1}}\argopen(##3\argclose)}}}{\mathinner{#2\IfNoValueTF{##1}{}{^{##1}}##2} \IfNoValueTF{##3}{}{(##3)}}}}
\makediff{mD}{\mathcal D}
\DeclareMathOperator{\T}{T}

\mathdef{\U}{\operatorname{U}}
\mathdef{\P}{\operatorname{P}}

\newcommand{\particle}[1]{\text{#1}}
\newcommand{\strange}{\particle{s}}

\newcommand{\up}{\particle{u}}

\newcommand{\epsilonEW}{\epsilon_{\EW}}

\newcommand{\fp}{f_0}

\usepackage{tocloft}

\AtBeginEnvironment{tabular}{\tlstyle}

\let\oldfigure\figure
\def\figure{\oldfigure\small}
\let\oldtable\table
\def\table{\oldtable\small}

\makeatletter
\def\deltabar{{\mathchar '26\mkern -9mu\delta}}
\def\custom#1{{\hbox{$\left#1\vbox to30\p@{}\right.\n@space$}}}
\makeatother

\graphicspath{{./figures/}}
\bibliography{bibliography}

\title{%
Factorizing Hidden Particle Production Rates}

\author[Bern]{Philipp Klose\footnote{\texttt{pklose@itp.unibe.ch}}} 

\affiliation[Bern]{Institut für theoretische Physik, Universität Bern}

\date{\today}

\begin{document}

\maketitle 

\begin{abstract}
A method is proposed to streamline the computation of hidden particle production rates by factorizing them into
i) a model-independent SM contribution, and ii) a observable-independent hidden sector contribution.
The SM contribution can be computed once for each observable and re-used for a wide array of hidden sector models,
while the hidden sector contribution can be computed once for each model, and re-used for a wide array of observables.
The SM contribution also facilitates extracting model independent constraints on hidden particle production.
The method is compatible with \EFT and simplified model approaches.
It is illustrated by factorizing the rate of charged kaon decays into a charged lepton and a number of hidden particles,
and a single form factor $F_{\ell}$ is found to parametrize the impact of general hidden sectors.
We derive model-independent constraints for the form factor $F_e$ that governs decays into positrons and hidden particles.
\end{abstract}

\tableofcontents
\clearpage

\section{Introduction}

Since the \SM of particle physics is known to be incomplete, the search for physics \BSM, and in particular for new particles, is an integral part of modern high energy physics.
Focusing on searches at collider and fixed target experiments, the phenomenology of candidate particles crucially depends on whether or not they are light enough to be produced at current generation experiments.
Light new particles are more strongly constrained, and they can only couple very feebly to the \SM.

In recent years, a significant program of searches for such feebly interacting new particles has begun to emerge \cite{Alekhin:2015byh, Beacham:2019nyx,Lanfranchi:2020crw,Agrawal:2021dbo}.
Relevant constraints can be obtained from high intensity data sets of \CMS \cite{Sirunyan:2018mgs, Sirunyan:2019sgo, Mukherjee:2019anz} and \ATLAS \cite{Aad:2020cje},
from flavour physics experiments such as \LHCb \cite{Aaij:2020iew, Aaij:2017rft, Aaij:2018mea, Aaij:2019bvg, Aaij:2020ikh, Aaij:2020ovh, Borsato:2021aum} or Belle-II \cite{Belle-II:2018jsg,Liang:2021kgw,Ferber:2022rsf}, and from high luminosity runs of the \LHC \cite{CidVidal:2018eel}.
There is also a number of ongoing and proposed searches at low-energy fixed target experiments such
\NA62 \cite{NA62:2017rwk,CortinaGil:2017mqf,CortinaGil:2018fkc,Drewes:2018gkc,CortinaGil:2019nuo,CortinaGil:2020fcx,NA62:2020mcv},
\KOTO \cite{Ahn:2018mvc,KOTO:2018dsc,Gori:2020xvq}, SeaQuest \cite{Aidala:2017ofy,Berlin:2018pwi}, or \SHiP \cite{Alekhin:2015byh},
and further complimentary constraints can be obtained from searches at dedicated \LLP experiments
\cite{Alimena:2019zri} such as \MATHUSLA \cite{Chou:2016lxi,Curtin:2018mvb}, \FASER \cite{Feng:2017uoz,Ariga:2018uku} and \CODEXb \cite{Gligorov:2017nwh}.

In spite of these searches, numerous viable hidden sector models still predict a large variety of hidden particle candidates.
Examples include \ALPs \cite{Peccei:1977ur, Wilczek:1977pj, Weinberg:1977ma, Peccei:1977hh,Lazarides:1985bj, Derendinger:1985cv, Giddings:1987cg,Alexander:2016aln},
\HNLs \cite{Drewes:2013gca, Alekhin:2015byh, Agrawal:2021dbo}, and new vector Bosons \cite{Alexander:2016aln,NA64:2018lsq,Fabbrichesi:2020wbt}.
Since standard perturbative methods of computing predictions for the relevant observables (i.e. hidden particle production, scattering, and decay rates)
depend on detailed knowledge of the number and the properties of the hidden fields that are presumed to exist as well as their interactions,
this poses a challenge for the model independent interpretation of the experimental constraints.

In recent years, \EFTs such as \SMEFT \cite{Buchmuller:1985jz, Grzadkowski:2010es} and \HEFT \cite{Feruglio:1992wf,Burgess:1999ha,Barbieri:2007bh,Grinstein:2007iv}
have become a standard tool for extracting model independent constraints on the \SM coupling to \emph{heavy} new particles \cite{Brivio:2017vri,Ellis:2018gqa, AguilarSaavedra:2018nen, Slade:2019bjo,Dong:2022mcv}.
Due to this success, some effort has been put towards constructing \EFTs that also account for light new particles.
For instance, there are a number of \EFTs that couple the \SM to specific candidate particles such as \ALPs \cite{Brivio:2017ije}, \HNLs \cite{Liao:2016qyd,Li:2021tsq}, or dark photons \cite{Barducci:2021egn}, as well as generic dark matter candidates \cite{Duch:2014xda,DeSimone:2016fbz,Contino:2020tix,Aebischer:2022wnl}.
\EFTs are also commonly used to describe non-relativistic interactions between the \SM and dark matter candidates \cite{Fitzpatrick:2012ix,Cirigliano:2012pq,DelNobile:2013sia,Hoferichter:2015ipa,Hoferichter:2016nvd,Bishara:2016hek,Bishara:2017pfq,Hoferichter:2018acd,Criado:2021trs}.
Finally, there has been significant work to create a comprehensive framework for constructing \PETs
that systematically extend \EFTs of the \SM by coupling them to generic hidden particles while making only a minimal number of assumptions \cite{Arina:2021nqi}.
Another approach for extracting model independent bounds consists in constructing so-called ``simplified models'', which are designed to capture certain features of realistic \SM extensions in a minimalist and therefore more generic setup.
Simplified models have become popular \eg within the context of searches for particles that are on the cusp of being collider accessible \cite{Alwall:2008ag,LHCNewPhysicsWorkingGroup:2011mji} and dark matter candidates \cite{Abdallah:2015ter,DeSimone:2016fbz}.
However, while both \EFTs and simplified models are useful for studying generic features of hidden sectors,
it is not always straightforward to translate the resulting constraints into hard constraints that are applicable to realistic \SM extensions.
\\

In this work, we focus on the computation of hidden particle production rates.
We argue that some of the challenges associated with establishing model independent constraints can be ameliorated by factorizing these production rates into a product of
i) model-independent ``reduced matrix elements'' that depend only on \SM interactions and
ii) observable-independent ``hidden current correlation matrices'' that capture the impact of general hidden sectors.
As we will show, this factorization is completely generic and relies only on a minimal set of assumptions.

Using the reduced matrix elements, it is possible to derive model-independent ``master formulae'' for a wide range of observables that parametrize the impact of general hidden sectors via a number of generic form factors.
The master formulae can be fitted to experimental data in order to extract model-independent constraints on the form factors.
In addition, it is possible to use the hidden current correlation matrices in order to translate the form-factor constraints into more specific constraints on the individual parameters of a given hidden sector model.
One major advantage of this two-step procedure is that it divides the workload required for computing hidden particle production rates into two largely independent packages:
On the one hand, the hidden current correlation matrices depend only on hidden sector physics and can be computed without having to account for the intricacies of \SM model physics or collider phenomenology.
On the other hand, the master-formulae depend only on \SM physics, and can be computed without having to specify the feature of the hidden sector.
This division also makes it possible to mix and match results in a more transparent and systematic way while minimizing the need for re-doing computations in order to adapt a given result to a new model.

In order to make use of the factorization procedure, it is necessary to supply a comprehensive list of relevant portal operators that can mediate a given production process.
In principle, any \EFT that accounts for the presence of light hidden particles can be used to provide such a list.
However, the \PET framework is particularly well suited for this task, since it makes only minimal assumptions about the symmetries obeyed by the portal interactions and about the internal structure of the hidden sector,
while also accounting for the presence of any additional heavy new particles.
When combined with an approriately constructed list of portal operators, the factorization approach is a powerful tool for establishing model independent constraints on hidden sectors.
It effectively extends the \EFT approach by providing a description that accounts for both light \emph{and} heavy new particles.
\\

The remainder of this paper is structured as follows:
In \cref{sec:inclusive production rates}, we provide a short proof of the factorization rule, and derive the general recipes for computing the reduced matrix elements and correlation matrices.
In \cref{sec:example computation}, we illustrate the procedure by considering the example of charged Kaon decays $K^+ \to \ell^+ X$ into a charged lepton and a number of hidden sector particles.
To do so, we first compute a model-independent master amplitude that encodes the impact of generic hidden sectors using a single form factor $F_\ell$, and then show how to compute
this form factor for an example model that couples the \SM to a number of hidden Fermions.
We also extract model-independent upper bounds on the form factor $F_e$ that governs charged kaon decays $K^+ \to e^+ X$ into a positron and a number of hidden particles.
\Cref{sec:conclusion} concludes the paper.

\section{Inclusive production rates}
\label{sec:inclusive production rates}

In this section, we demonstrate the factorization of inclusive hidden sector production rates into a product of reduced matrix elements $M_d$ and hidden sector correlation matrices $J^{de}$.
Although the proof is not complicated, it also serves as a derivation of the diagrammatic expressions for both quantities.
In the interest of full generality, we consider a generic theory
that is composed of a visible sector $A$ and a hidden sector $B$,
\begin{align} \label[lag]{eq:portal action}
\mathcal L &= \mathcal L_A + \mathcal L_B + \mathcal L_\text{portal} \ , &
\mathcal L_\text{portal} &= \sum_d \epsilon \mathcal A_d \, \mathcal B^d \ .
\end{align}
The two sectors are linked by a number of weak portal interactions whose strength is measured by a small parameter $\epsilon$.
Each portal operator is the product of a local operator $\mathcal A_d = \mathcal A_d[\{\phi_a\}](x)$ constructed from a collection of visible fields $\{\phi_a\}$
and a conjugated local operator $\mathcal B^d = \mathcal B^d[\{\xi_c\}](x)$ constructed from a collection of hidden fields $\xi_b$.

We are interested in computing inclusive rates for the production of hidden particles in experimental setups where the hidden particles cannot be observed directly.
In this case, only the number and the properties of the visible particles in the final state are known, and the resulting transition rates involve a sum over an infinite number of viable final states,
\begin{align}
\label{eq:inclusive transition rate}
\abs{M}^2 &= \sum_{k=1}^{\infty} \abs{M_k}^2 \ , &
\abs{M_k(\mathcal K \to \mathcal P)}^2 &= \int \hspace{-3pt} \text{d}^3_k \mathcal Q \ \deltabar^4 (K - P - Q) \left| \mathcal M ( \mathcal K \to \mathcal P ; \mathcal Q ) \right|^2 \ , &
\end{align}
where the index $k$ denotes the total number of hidden particles in the final state.
The matrix elements
\begin{align}
\label{eq:matrix element}
\deltabar^4 (K-P-Q) \mathcal M (\mathcal K \to \mathcal P; \mathcal Q) &= \langle \mathcal P; \mathcal Q | \i \T | \mathcal K \rangle \ , &
\deltabar^4(K) &= (2\pi)^4 \delta^4 (K)
\end{align}
encode the likelihood of transitioning from an initial state
\begin{align}
|\mathcal K \rangle &= | \bm k_1^{s_1} {\dots \ } \bm k_n^{s_n} \rangle \ ,
\end{align}
that consists of $n$ visible particles $s_i$ with four-momenta $k_i = (\kappa_i, \mathbf k_i)$ into a given final state
\begin{align}
\langle \mathcal P; \mathcal Q | &= \langle \mathcal P | \otimes \langle \mathcal Q | \ , &
\langle \mathcal P | &= \langle \bm p_1^{t_1} {\dots } \bm p_m^{t_m}| \ , &
\langle \mathcal Q | &= \langle \bm q_1^{r_1} {\dots } \bm q_k^{r_k} | \ .
\end{align}
that consists of $m$ visible particles $t_j$ with four-momenta $p_j = (\pi_j, \mathbf p_j)$ and $k$ hidden particles $r_l$ with four-momenta $q_l = (\omega_l, \mathbf q_l)$.
Here and in the following, we use the multi-indices $s = (s_1, \dots, s_n)$, $t = (t_1, \dots, t_m)$, and $r = (r_1, \dots, r_k)$ to collectively denote the species and the helicity of each particle.
The integration measure
\begin{align}
\text{d}^3_k \mathcal Q &= \prod_{l=1}^k \frac{\text{d}^3 \bm q_l}{(2\pi)^3} \frac1{2\omega_l} \sum_{r_l} \ ,
\end{align}
includes a sum over both the species and the helicity of each hidden particle as well as an integral over its associated phase space.
Finally, the sums
\begin{align}
K &= \sum_{i} k_i \ , &
P &= \sum_j p_j \ , &
Q &= \sum_l q_l = K - P
\end{align}
denote the total momentum of the visible particles in the initial and final states as well as the corresponding missing momentum.

Most of the relevant production modes for hidden particles involve either decays or scatterings of visible particles.
The general sum \eqref{eq:inclusive transition rate} determines the overall rates for hidden particle production in both of these cases, yielding
\begin{align}
\text{d} \mathrm \Gamma &= \frac1{2 \kappa_1} \abs{M(k_1 \to \mathcal P)}^2 \prod_{j=1}^m \frac{\text{d}^3 \bm p_j}{(2\pi)^3} \frac1{2\pi_j} \ ,
&\text{d} \sigma &= \frac1{4 \kappa_1 \kappa_2 v_{12}}  \abs{M( k_1, k_2 \to \mathcal P)}^2 \prod_{j=1}^m \frac{\text{d}^3 \bm p_j}{(2\pi)^3} \frac1{2\pi_j} \ ,
\end{align}
where $v_{12} = \abs{v_1 - v_2}$ is the relative velocity of the two initial state particles in the scattering process.
Our aim is to show that the inclusive rate $M(\mathcal K \to \mathcal P)$ factorizes according to
\begin{align}
\label{eq:inclusive rate factorization}
\abs{M (\mathcal K \to \mathcal P)}^2 &= \epsilon^2 M_d M_e^\dagger J^{de}+ \order{\epsilon^3} \ ,
\end{align}
where the reduced matrix elements $M_d$ encode the dynamics of the visible sector while the hidden current correlation matrix $J^{de}$ encodes the impact of hidden sectors.
We note that, although \cref{eq:inclusive rate factorization} is completely general, its usefulness crucially depends on the size of the small parameter $\epsilon$.
However, this is of no concern within the context of hidden sector searches, since higher order corrections are almost always negligible due to the required smallness of the portal coupling.
In this case, the factorization \cref{eq:inclusive rate factorization} becomes a very good approximation.

\subsection{Factorization}

The factorization of the inclusive rate \eqref{eq:inclusive rate factorization} is a corollary of an equivalent factorization of the time-ordered, connected correlation functions
\begin{align}
\hat G_{abc} (\mathcal X, \mathcal Y, \mathcal Z) = \langle 0 | \T \{ \hat \Phi_a (\mathcal X) \hat \Phi^\dagger_b(\mathcal Y) \hat \Xi_{c}^\dagger (\mathcal Z) \} | 0 \rangle_{\text{conn.}}
\end{align}
 that capture the dynamics of hidden particle production.
Here and in the following, the average $\langle 0 | \circ | 0 \rangle_{\text{conn.}}$ is defined to include only connected diagrams.
In order to avoid superfluous notational clutter, we have defined the multi-field operators
\begin{align}
\hat \Phi_a (\mathcal X) &= \prod_{i=1}^n \phi_{a_i} (x_i) \ , &
\hat \Xi_a (\mathcal X) &= \prod_{l=1}^k \xi_{a_i} (x_i) \ , &
a &= (a_1, \dots , a_n) \ , &
\mathcal X &= (x_1, \dots, x_n) \ .
\end{align}
We also define the corresponding momentum-space operators
\begin{align}
\Phi_a( \mathcal K) &= \int \text{d}^4_n \mathcal X e^{\i \mathcal X \mathcal K} \hat \Phi_a(\mathcal X) \ , &
\text{d}^4_n \mathcal X &= \prod_{i=1}^n \text{d}^4 x_i \ , &
\mathcal X \mathcal K &= \sum_{i=1}^n x_i k_i \ .
\end{align}
Expanding the path-integral of the full theory to leading order in $\epsilon$, one immediately finds that the above correlation functions factorize according to
\begin{align}
\hat G_{abc} &= - \i \epsilon \int \text{d}^4 x \ \hat G_{ab;d} \hat F_{c;d} + \order{\epsilon^2} \ , &
\hat G_{ab;d} &= \i \langle 0 | \T \{ \hat \Phi_a (\mathcal X) \hat \Phi^\dagger_b(\mathcal Y) \mathcal A_d(x) \} | 0 \rangle_{\text{conn.}}^{\epsilon \to 0} \ ,
\end{align}
where the reduced correlation function $\hat G_{ab;d}$ only depends on physics of the visible sector, while the form factor
\begin{align}
\hat F_{c;d} &= \i \langle 0 | \T \{ \hat \Xi_{c}^\dagger (\mathcal Z) \mathcal B^d (x) \} | 0 \rangle_{\text{conn.}}^{\epsilon \to 0}
\end{align}
encodes the impact of hidden sectors.
Moving on to momentum space, this gives
\begin{equation}
\label{eq:greens function factorization}
G_{abc}
= \langle 0 | \T \{ \Phi_a (\mathcal K) \Phi^\dagger_b(\mathcal P) \Xi_{c}^\dagger (\mathcal Q) \} | 0 \rangle_{\text{conn.}}
= - \i \epsilon \, G_{ab;d} \left(\mathcal K, \mathcal P \right) \mathcal F_{c;d} \left(\mathcal Q \right) + \order{\epsilon^2} \ ,
\end{equation}
where the momentum space versions of the reduced Greens functions and the hidden sector form factors are
\begin{subequations}\label{eq:momentum space}
\begin{align}
\label{eq:Gd}
\deltabar^4 (K-P-q) G_{ab;d} &=
\int \text{d}^4 x \, \text{d}^4_n \mathcal X \, \text{d}^4_m \mathcal Y \, e^{\i \left(\mathcal K \mathcal X - \mathcal P \mathcal Y - q x \right)} \hat G_{ab;d} \ , \\
\label{eq:Fd}
\deltabar^4 (q-Q) F_{c;d} &=
\int \text{d}^4 x \, \text{d}^4_k \mathcal Z \, e^{\i \left(q x - \mathcal Q \mathcal Z \right)} \hat F_{c;d} \ .
\end{align}\end{subequations}
In order to translate the factorization \cref{eq:greens function factorization} into a statement about the inclusive transition rate \eqref{eq:inclusive transition rate},
we apply the Lehmann-Symanzik-Zimmermann (LSZ) reduction formula to \eqref{eq:matrix element}.
Using the multi-field notation, this gives
\begin{align}
\label{eq:LSZ reduction}
\i \mathcal M (\mathcal K \to \mathcal P; \mathcal Q) &= \mathcal E^a \overline{\mathcal E}^b \overline{\mathcal E}^c G_{abc} \ , &
\mathcal E^a &= \prod_i Z_{a_i}^{-\nicefrac12} \epsilon^{b_i} G_{b_i a_i}^{-1} \ , &
\overline{\mathcal E}^a &= \prod_i Z_{a_i}^{-\nicefrac12} \overline \epsilon^{b_i} G_{b_i a_i}^{-1} \ ,
\end{align}
where the amputating factors $\mathcal E^a = \mathcal E^a (s, \mathcal K) $ and $\overline{\mathcal E}^a = \overline{\mathcal E}^a (s, \mathcal K)$ consist of the standard wave-function renormalization factors $Z_{a_i}$,
the initial and final state polarization vectors $\epsilon^{a_i} = \epsilon^{a_i}(s_i, k_i)$ and $\overline \epsilon^{a_i} = \overline \epsilon^{a_i} (s_i, k_i)$,
and the inverse propagators $G^{-1}_{b_i a_i}(k_i)$.%
\footnote{
Recall that the multi-indices $s = (s_1,\dots, s_n)$ collectively denote the species and the helicity of each \emph{particle}, while the multi-indices $a = (a_1, \dots, a_n)$ collectively denote the type of each \emph{field}.
Each $a_i$ runs over all available fields present in the theory, and the polarization vectors are defined such that $\epsilon^{a_i} (s_i, k_i) = 0$ in cases where $s_i$ denotes a particle that is not produced by the field associated with the index $a_i$.}
With these definitions in place, inserting \cref{eq:greens function factorization,eq:LSZ reduction} into \cref{eq:inclusive transition rate} finally yields \cref{eq:inclusive rate factorization},
where
\begin{align}
\label{eq:Md Jd}
\i M_d &= \mathcal E^a \overline{\mathcal E}^b G_{ab;d} \ , &
J_{de} &= \sum_{k=1}^{\infty} \int \hspace{-3pt} \mD_k \mathcal Q \ \deltabar^4(q-Q) J_d J_e^\dagger \ , &
\i J_d &= \i J_d (\mathcal Q, r) = \overline{\mathcal E}^c F_{c;d} \ .
\end{align}
The external current correlation matrix $J_{de}$ fully encodes the impact of hidden sectors.
Since the integration $\sum_k \int \mD_k \mathcal Q$ in the definition of $J_{de}$ includes a sum over the species and the helicities of the hidden particles in the final state, the correlation matrix 
\begin{align}
J_{de} = \sum_{k=1}^{\infty} \sum_{r} J_{de}^r
\end{align}
can be written as an infinite sum of terms $J_{de}^r = J_{de}^r(Q)$ that capture the contribution associated with each individual final state.
Notice also that $M_d$ and $J_d$ (and with them $J_{de}$) can, depending on the precise structure of the corresponding portal operator, carry free Lorentz and spinor indices.
For example, if the theory contains a portal operator $\mathcal A_\mu \mathcal B^\mu = \psi \sigma_\mu \psi^\dagger v^\mu$
that couples a pair of visible Fermions $\mathcal A_\mu = \psi \sigma_\mu \psi^\dagger$ to a hidden vector particle $\mathcal B^\mu = v^\mu$,
then the corresponding reduced matrix element $M^\mu$ and hidden currents $J_\mu$ carry free Lorentz indices that are to be contracted with each other. 

\subsection{Feynman rules}

The expressions \eqref{eq:momentum space,eq:Md Jd} define series of Feynman diagrams that can be used to compute the reduced matrix elements $M_d$
as well as the hidden currents $J_d (r,\mathcal Q)$ associated with each viable hidden sector final state.
\begin{figure}
\centering
\begin{subfigure}[b]{0.45\textwidth}
\begin{equation*}
\i M_d (\mathcal K \to \mathcal P) = \quad \mathcal K \ \custom\{ \raisebox{-0.25\height}{\includegraphics[width=.5\textwidth, trim = 20 20 20 20, clip]{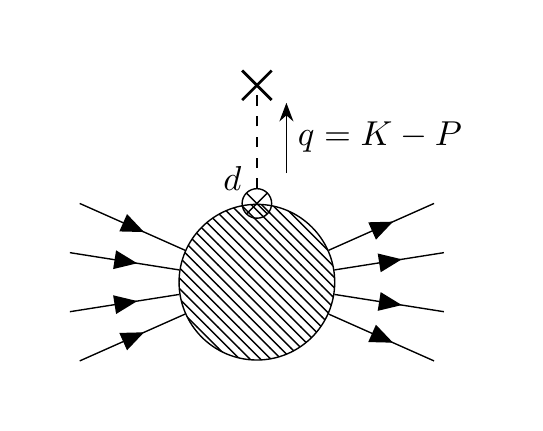}} \custom\} \ \mathcal P
\end{equation*}
\caption{\label{subfig:reduced matrix elements}Reduced matrix elements}
\end{subfigure}
\hfill
\begin{subfigure}[b]{0.45\textwidth}
\begin{equation*}
\i J_d (\mathcal Q) = \quad \vcenter{\hbox{\includegraphics[width=.5\textwidth, trim = 20 20 20 20, clip]{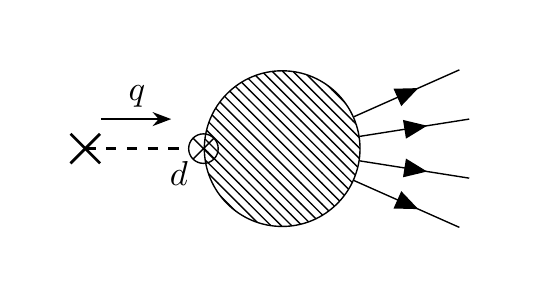}}} \custom\} \ \mathcal Q
\end{equation*}
\caption{\label{subfig:hidden currents}Hidden currents}
\end{subfigure}
\caption{\label{fig:diagrams} Diagrams for computing the reduced matrix elements $M_d$ and the hidden currents $J_d$.
Both are given as the sum of all available connected and amputated Feynman diagrams with the appropriate number and kind of particles in the initial and final states.
The crossed dot denotes the required portal vertex, and the dashed line denotes the relevant missing momentum in- and outflow.
Aside from this portal vertex, diagrams for $M_d$ diagrams only contain vertices and propagators associated with visible fields and interactions,
while diagrams for $J_d$ only contain vertices and propagators associated with hidden fields and interactions. 
}
\end{figure}

\Cref{fig:diagrams} shows the diagrammatic expansions for both $M_d$ and $J^d$.
As in the case standard S-matrix element computations, they are given as a sum of all available connected and amputated Feynman diagrams with the appropriate number and kind of particles in the initial and final states,
where $M_d$ diagrams only contain vertices and propagators associated with visible fields and interactions while $J_d$ diagrams only contain vertices and propagators associated with hidden fields and interactions. 
The main difference compared to the recipe for standard S-matrix elements is that all relevant diagrams have to contain exactly one portal vertex
that is constructed from either the ``visible'' part $\mathcal A_d$ (in the case of $M_d$) or the ``hidden'' part $\mathcal B^d$ (in the case of $J_d$) of the corresponding portal operator.

The rules for constructing these portal vertices are largely the same as those for constructing standard vertices.
(i.e. symmetrize the operator under exchange of identical fields, go to momentum space, strip away all fields, and multiply by an overall factor of $\i$.)
However, there are two differences:
First, and in complete analogy to the composite operators $\mathcal A_d$ and $\mathcal B^d$, the portal vertices may carry free Lorentz and spinor indices, which result in $M_d$ and $J^d$ likewise carrying such free indices.
Second, both portal vertices do not conserve four-momentum in the strict sense.
Rather, the appropriate sum of all ingoing and outgoing momenta has to be equal to some missing momentum $q$.
This is depicted symbolically using a dashed line (\cf \cref{fig:feynman rules}).
The missing momentum is outflowing for $M_d$, so that $q = K - P$, and it is inflowing for $J_d$, so that $q = Q$.

\section{Illustrative example: Inclusive $K^+ \to \ell^+$ decays}
\label{sec:example computation}

To illustrate the factorization procedure as well as the computation of the reduced matrix elements and hidden currents,
we consider the production of hidden particles in inclusive $K^+ \to \ell^+ X$ decays, where $X$ denotes any collection of hidden particles.
There already exists a model-independent master-formula for the production rate of generic spin \textfrac12 particles in this type of charged kaon decay \cite{Arina:2021nqi},
and the present computation improves this result by accounting for the production of hidden particles with different spins as well as the production of multiple hidden particles via the same decay process.

\subsection{Reduced matrix elements} \label{sec:reduced amplitudes}

\begin{figure}
\centering
\hfill
\begin{subfigure}[b]{0.32\textwidth}
\begin{equation*}
\vcenter{\hbox{\includegraphics[width=.525\textwidth, trim = 20 20 20 20, clip]{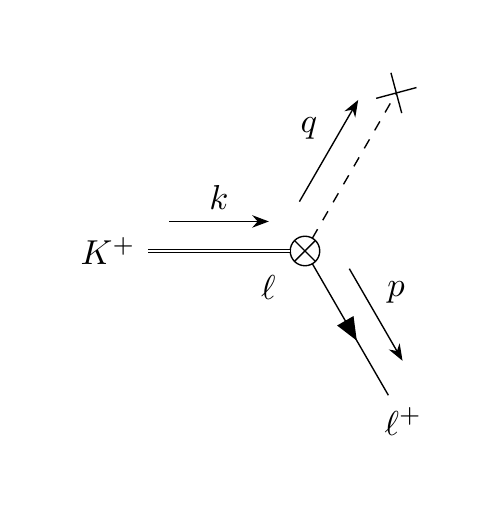}}}
\ \ = \frac{V_{\up\strange} \fp}{v^2} k_\mu \overline \sigma^\mu
\end{equation*}
\caption{\label{subfig:kaon vertex}Kaon portal vertex}
\end{subfigure}
\hfill
\begin{subfigure}[b]{0.32\textwidth}
\begin{equation*}
\vphantom{\vcenter{\hbox{\includegraphics[width=.525\textwidth, trim = 20 20 20 20, clip]{kaon_portal_vertex}}}}
\vcenter{\hbox{\includegraphics[width=.7\textwidth, trim = 20 20 20 20, clip]{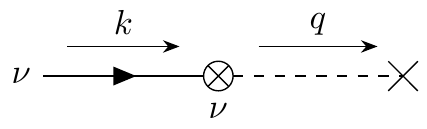}}}
\ \ = \i v
\end{equation*}
\caption{\label{subfig:neutrino vertex}Neutrino portal vertex}
\end{subfigure}
\hfill
\begin{subfigure}[b]{0.32\textwidth}
\begin{equation*}
\vphantom{\vcenter{\hbox{\includegraphics[width=.525\textwidth, trim = 20 20 20 20, clip]{kaon_portal_vertex}}}}
\vcenter{\hbox{\includegraphics[width=.7\textwidth, trim = 20 20 20 20, clip]{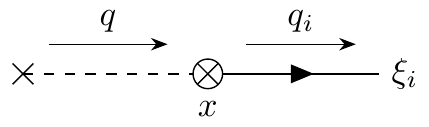}}}
\ \ = \i c_{x i}
\end{equation*}
\caption{\label{subfig:hidden vertex}Hidden fermion portal vertex}
\end{subfigure}
\hfill
\caption{\label{fig:feynman rules} Feynman rules for portal vertices that appear in the example computation of inclusive $K^+ \to \ell^+$ decay rates.
The first vertex \cref{subfig:kaon vertex} determines the size of the reduced matrix element $M^\ell$, while the second vertex \cref{subfig:neutrino vertex} determines the size of $M^\nu$.
Both contribute at the same order because $M^\nu$ is also suppressed by the smallness of the required additional Fermi-theory vertex associated with $K^+ \to \ell^+ \nu$ decays.
The third vertex \cref{subfig:hidden vertex} determines the size of the hidden currents $J^\ell$ and $J^\nu$, and is used in \cref{subsec:hidden current matrices}.}
\end{figure}

The primary input required to factorize inclusive matrix elements according to \eqref{eq:inclusive rate factorization} is an exhaustive list of all relevant portal operators.
Here, we use the list compiled in \cite{Arina:2021nqi}, which encompasses all leading operators that couple the \SM to a single hidden particle of spin $0$, $\nicefrac12$, or $1$ at the strong scale.
The sector of the portal Lagrangian that is relevant for $K^+ \to \ell^+ X$ decays is
\begin{align} \label[lag]{eq:LE portal Lagrangian}
\mathcal L_\text{portal} &\supset - v \nu \Xi_\nu + \frac{V_{\up \strange}}{v^2} Q_\mu \Xi_\ell^\dagger \overline \sigma^\mu \ell  \ , &
Q_\mu &= s^\dagger \overline \sigma_\mu u \ ,
\end{align}
where $u$ and $s$ are the up- and strange quark fields, $\ell = e, \mu$ is a charged lepton field, and $\nu$ is the corresponding \SM neutrino field.
Following two-component notation of \cite{Dreiner:2008tw}, the \SM Fermion fields are taken to be left-handed Weyl spinors.
$ v = \unit[174.10358 \pm 0.00004]{GeV}$ is the vacuum expectation value of the Higgs Boson and $\abs{V_{\up\strange}} = 0.2252 \pm 0.0008$ is the $\up$-$\strange$ element of the \CKM matrix \cite{ParticleDataGroup:2020ssz}.
The small parameter $\epsilon$ has been re-absorbed into the operators $\Xi_\nu, \Xi_\ell$, which now take on the role of the generic local operators $\epsilon \, \mathcal B^d$.
It is not necessary to specify their detailed shape in order to compute the reduced matrix elements.
However, we note that, although the master formula that results from using the portal \cref{eq:LE portal Lagrangian} is largely model-independent,
the factorization procedure outlined here can also be used to obtain an even more general master formula by including additional sub-leading portal operators.
A collection of the relevant interactions is given e.g. in \cite{Liao:2016qyd,Li:2021tsq}.

Given the portal \cref{eq:LE portal Lagrangian}, the resulting coupling of hidden sectors to the pseudoscalar mesons is captured by the portal \cPT action constructed in \cite{Arina:2021nqi}.
We work at leading order in the \cPT power counting, at tree-level, and neglect electromagnetic corrections.
At this level of accuracy, the coupling of the operator $\Xi_\ell$ to charged kaons can be obtained by replacing the quark bilinears according to \cite{Gasser:1984gg,ParticleDataGroup:2020ssz,Arina:2021nqi}
\begin{align}
Q_\mu &\to \fp \partial_\mu K^+ \ , &
\fp &= \unit[77.85 \pm 0.15]{MeV} \ ,
\end{align}
where $\fp$ is the kaon decay constant.
This replacement yields the ``visible'' portal vertex depicted in \cref{subfig:kaon vertex}, while neutrino portal operator in \eqref{eq:LE portal Lagrangian} yields the additional vertex depicted in \cref{subfig:neutrino vertex}.

In general, there is one reduced matrix element for each portal operator.
Since there are two relevant portal operators, there are also two reduced matrix elements $M_\ell$ and $M_\nu$.
Using the Feynman rules for two-component spinors given in \cite{Dreiner:2008tw,Martin:2012us}, one finds the leading order expressions%
\footnote{The leading contribution to $M_\ell$ is generated by the tree-level diagram that contains only the portal vertex in \cref{subfig:kaon vertex},
and the leading contribution to $M_\nu$ is generated by the tree-level diagram that contains the \SM three-point vertex that mediates $K^+ \to \ell^+ \nu_\ell$ decays
as well as the portal vertex in \cref{subfig:neutrino vertex}}
\begin{align}
M_\nu^\alpha &= \i \frac{\fp V_{\up\strange}}{v q^2} (\sigma_\nu \overline \sigma_\mu)^{\alpha \beta} y_\beta^\ell(p,t) q^\nu k^\mu \ , &
M_\ell^{\dot \alpha} &= - \i \frac{\fp V_{\up\strange}}{v^2} (\overline \sigma_\mu)^{\dot \alpha \beta} y_\beta^\ell(p,t) k^\mu \ ,
\end{align}
where $y^\ell$ is the two-component polarization vector of the final state lepton,
$k = (\kappa, \bm k)$ is the kaon four-momentum, $p = (\pi, \bm p)$ is the lepton four-momentum, and $q = k-p$ is the missing momentum.
The free spinor indices $\alpha$ and $\dot \alpha$ are to be contracted with the external current correlation matrices.	
Summing over the available $\ell^+$ spin polarizations, the resulting inclusive rate for $K^+ \to \ell^+$ decays is
\begin{subequations}\begin{align}
\sum_{t} \abs{M (k \to p)}^2
&= \sum_t \left( M_\ell^{\dot\alpha} M_\ell^{\dagger\beta} J_{\beta\dot\alpha}^{\ell\ell} + M_\nu^\alpha M_\nu^{\dagger\dot\beta} J_{\dot\beta\alpha}^{\nu\nu} + 2 \Re M_\nu^\alpha M_\ell^{\dagger\beta} J_{\beta\alpha}^{\ell \nu} \right) \\
\label{eq:K to ell master formula}
&= \abs{\frac{\fp V_{\up\strange}}{v^2}}^2 \frac{2 (pk) (qk) - (pq) k^2}{k^2} \ F_\ell(x_q) \ .
\end{align}\end{subequations}
The impact of hidden sector contributions is encoded by the form factor
\begin{align}
\label{eq:kaon form factor}
x_q F_\ell\left(x_q\right) &= \tr_D \left\{
q^\nu \overline \sigma_\nu J^{\ell\ell}
- 2 \frac{v}{q} \Re q J^{\ell\nu}
+ \frac{v^2}{q^2} q^\rho \sigma_\rho J^{\nu\nu} \right\} \ , &
x_q &= \frac{q^2}{m_K^2} \ ,
\end{align}
where $\tr_D$ is the trace taken with respect to the (un)dotted spinor indices.
Since the Levi-Civita tensor is used to raise and lower indices associated with the same chirality, $V^\alpha = \epsilon^{\alpha\beta} V_\beta$,
one has $\tr_D \left\{ J^{\ell a}\right\} = \epsilon^{\alpha\beta} J^{\ell a}_{\beta \alpha}$.
This gives the partial decay width
\begin{align}
\label{eq:model independent decay width}
\frac{\text{d}\mathrm\Gamma_\ell}{\text{d}x_q} &= 4 \pi m_K \abs{V_{\up\strange}}^2 \left( \epsilonEW \frac{m_K}{4\pi\fp} \right)^2 \rho \ \frac{F_\ell(x_q)}{2\pi} \ , &
\epsilonEW &= \frac{\fp^2}{v^2} \ ,
\end{align}
where
\begin{align}
\rho(x_\ell,x_q) &=\frac12 \left( x_\ell + x_q - (x_\ell - x_q)^2 \right) \sqrt{ \left( \frac{1 - x_\ell - x_q}2 \right)^2 - x_\ell x_q } \ , &
x_\ell &= \frac{m_\ell^2}{m_K^2}
\end{align}
is a phase-space factor, $m_K = \unit[493.636 \pm 0.007]{MeV}$ is the charged kaon mass, and $m_\ell$ is the mass of the relevant charged lepton \cite{ParticleDataGroup:2020ssz}.
Comparing \cref{eq:model independent decay width} with the known partial width for $K^+ \to \ell^+ \nu$ decays \cite{Cirigliano:2011ny}, one finally obtains the ratio of branching ratios
\begin{align}
\label{eq:branching ratio}
R_\ell(x_q) &= \frac{\flatfrac{\text{d} \mathcal B_\ell}{\text{d} x^2}}{\mathcal B(K^+ \to \ell^+ \nu)}
= \frac{ \flatfrac{\text{d}\mathrm \Gamma_\ell}{\text{d}x^2} }{\mathrm \Gamma(K^+ \to \ell^+ \nu)}
= \frac{\rho(x_\ell,x_q)}{\rho(x_\ell,0)} \frac{F_\ell(x_q)}{2\pi} \ .
\end{align}
Since the master formula \eqref{eq:model independent decay width} only depends on the overall shape of the portal \cref{eq:LE portal Lagrangian},
but not on the specifics of the portal interactions or on the internal structure of the hidden sector, formula \eqref{eq:branching ratio} is almost completely model independent.
In particular, it captures the production of an arbitrary number of hidden particles with arbitrary masses, spins, and interactions.
Since we have also allowed for the possibility of higher dimensional \SM and portal operators, formula \eqref{eq:branching ratio} even accounts for the presence of heavy new particles.

Formula \eqref{eq:branching ratio} can be used to extract model-independent constraints on the size of the form factors $F_\ell(x_q)$.
To illustrate how this can be done in practice, we re-interpret the analysis in \cite{NA62:2020mcv},
which uses a missing mass search to establish upper bounds for the branching ratio of charged kaon decays into a positron and a \HNL from NA62 data.
The \HNLs were assumed to decay outside the detector, and the search was hence directed at finding events with a single visible positron in the final state and some finite missing mass $q^2 = m_\text{miss}^2 = (k-p)^2$.
This setup is a special case of the generic setup that we have considered in this work, where the number and the type of visible particles in the final state is known, while any produced hidden particles are not observed directly.

The analysis in \cite{NA62:2020mcv} sampled a large number of missing masses in the range $\unit[122]{MeV} < m_\text{miss} < \unit[465]{MeV}$, searching for candidate events in a bin of width $2 \Delta m = \unit[0.3]{MeV}$ centered around the sampled missing mass.
For each bin, they extracted an upper bound on the corresponding branching ratio.
This upper bound is roughly constant over the entire mass range, and comes out to $\mathcal B(K^+ \to e^+ N) \lesssim 4 \cdot 10^{-9}$.
In the following, we interpret this number as a bound on the fractional branching ratio $\Delta \mathcal B_e = \Delta x_q \cdot \flatfrac{\text{d}\mathcal B_e}{\text{d}x_q}$,
which captures the likelihood of producing a positron and a collection of generic hidden particles with an aggregate missing mass in the range $\left[x_q - \Delta x_q, x_q + \Delta x_q \right]$, where $\Delta x_q = (\flatfrac{\Delta m}{m_K})^2$.
Using the known values of $\mathcal B(K^+ \to e^+ \nu_e) = (1.582 \pm 0.007)\cdot 10^{-5}$ and $x_e = \flatfrac{m_e^2}{m_K^2} \approx 1.07 \cdot 10^{-6}$ \cite{ParticleDataGroup:2020ssz},
this gives the constraint
\begin{align}
\label{eq:F bound}
\rho(x_e, x_q) \frac{\langle F_e \rangle 2 \Delta x_q}{2\pi} &\lesssim 7 \cdot 10^{-11} \ , &
2 \Delta x_q &\approx 4 \cdot 10^{-7} \ , &
0.06 &< x_q < 0.89 \ ,
\end{align}
where $\langle F_e \rangle$ is the average of $F_e(x_q)$ taken over the range $\left[x_q - \Delta x_q, x_q + \Delta x_q \right]$.
To understand how to interpret this constraint, recall that the form factor captures the production of an arbitrary number of hidden particles.
Generically, a contribution associated the production of a single hidden particle will include the delta distribution $\delta (q^2 - m_i^2)$, where $m_i$ is the mass of the relevant hidden particle, to ensure that only on-shell particles are produced.
In contrast, the phase-space integral $\int \text{d}_k Q$ in \cref{eq:Md Jd} is sufficient to smooth out the contributions associated with the production of two or more hidden particles.
Hence, the form factor is of the general shape
\begin{align}
\label{eq:form factor shape}
\frac{F_e (x_q)}{2\pi} &= \sum_i A_i \delta(x_q^2 - x_i^2) + B(x_q) \ , &
x_i &= \frac{m_i^2}{m_K^2} \ ,
\end{align}  
where the $A_i$ are amplitudes for single particle production and $B (X_q)$ is a smooth function that captures the production of multiple hidden particles.
Averaging \eqref{eq:form factor shape}, one finds
\begin{align}
\label{eq:F decomposition}
\frac{\langle F_e \rangle 2 \Delta x_q}{2\pi} &= \sum_i A_i \Theta\left( \Delta x_q - \abs{x_i - x_q} \right) + B(x_q) 2 \Delta x_q \ ,
\end{align}
where we have used that $\langle B \rangle \approx B$ for sufficiently small bins.
Combining this decomposition with the general expression \eqref{eq:F bound},
one now obtains separate constraints on the single-particle amplitudes $A_i$ and the multi-particle amplitude $B(x_q)$.
The single-particle amplitudes have to obey the model-independent constraint
\begin{align}
\label{eq:A bound}
\rho(x_e, x_i) A_i &\lesssim 7 \cdot 10^{-11} \ , &
0.06 &< x_i < 0.89 \ ,
\end{align}
while the constraint on the multi-particle amplitude is much less stringent,
\begin{align}
\label{eq:B bound}
\rho(x_e, x_q) B(q^2) &\lesssim 2 \cdot 10^{-4} \ , &
0.06 &< x_q < 0.89 \ .
\end{align}
This is to be expected, since a signal for the production of a single hidden particle would appear as a sharp peak that is concentrated into a single bin,
while a signal for the production of multiple particles would be spread over a whole range of viable missing masses.
While the peak is relatively easy to observe, and therefore constrain, it is more difficult to constrain the spread-out signal.

\subsection{Hidden current matrices}
\label{subsec:hidden current matrices}

In order to translate constraints on the form factors $F_\ell(x_q)$ into more specific constraints, it is necessary to plug in appropriate expressions for the hidden current correlation matrices $J^{de}$.
In this section we demonstrate how to compute general correlation matrices by considering an example case in which the \SM couples to a number of left-handed Weyl Fermions $\xi_i$.
Since both Dirac and Majorana Fermions can always be written as a combination of Weyl Fermions, this setup remains quite general.
The leading contributions to the portal operators are
\begin{align}
\label{eq:fermion currents}
\Xi_d &= \sum_i c_{d i} \xi_i \ , &
d &= \nu, \ell \ ,
\end{align}
where the constants $c_{\nu, \ell}$ are model-dependent coupling constants.
The corresponding ``hidden'' portal vertices are depicted in \cref{subfig:hidden vertex}.

In general, there is one collection of hidden currents $J^d = J^d (\mathcal Q, r)$ for each viable hidden sector final state $r = (r_1, \dots, r_k)$, and each of these collections contains one current for each available portal operator.
In the concrete case of $K^+ \to \ell^+ X$ decays this means that there are two hidden currents $J^\nu(\mathcal Q, r)$ and $J^\ell (\mathcal Q, r)$ for each final state.
Assuming that the hidden sector interactions are perturbatively small, the leading contributions are generated by diagrams without any hidden sector vertices.
Since the portal operators in \cref{eq:fermion currents} contain only a single field operator,
the only final states that are viable at this order of accuracy (i.e. neglecting hidden sector interactions) are those with a single hidden Fermion, and no additional hidden particles.
The two hidden currents that are associated with this type of final state are given as
\begin{align}
J_\alpha^{\nu}(q_1,r_1) &= c_{\nu i} y_\alpha^i (\bm q_1,r_1) \ , &
J_{\dot \alpha}^{\ell}(q_1,r_1) &= c_{\ell i} x_{\dot \alpha}^{i\dagger} (\bm q_1,r_1) \ ,
\end{align}
where index $i = i(r_1)$ denotes $\xi_i$ that creates the single-particle final state associated with $r_1$.
Evaluating the spin sums and the phase-space integration, one obtains the correlation matrices
\begin{subequations}
\label{eq:correlation matrices}
\begin{align}
J_{\dot\beta\alpha}^{\nu\nu} &= \sum_i \frac{c^\dagger_{\nu i} c_{\nu i}}{2\omega_i} (q_i^\mu\overline \sigma_\mu)_{\dot\beta\alpha} 2\pi \delta (q_0 - \omega_i) \ , &
J_{\beta\alpha}^{\ell \nu} &= \sum_i \frac{c_{\ell i}^\dagger c_{\nu i}}{2\omega_i} m_i \epsilon_{\beta\alpha} 2\pi \delta (q_0 - \omega_i) \ , \\
J_{\beta\dot\alpha}^{\ell\ell} &= \sum_i \frac{c_{\ell i}^\dagger c_{\ell i}}{2\omega_i} (q_i^\mu\sigma_\mu)_{\beta\dot\alpha} 2\pi \delta (q_0 -\omega_i) \ ,
\end{align}\end{subequations}
where $m_i$ is the mass of the hidden Fermion $\xi_i$.
Inserting these matrices into the form factor \cref{eq:kaon form factor}, one finds
\begin{align}
\frac{F_\ell(x_q)}{2\pi} &= \sum_i U_i^2 \Theta(q_0) \delta (x_q^2 - x_i^2) \ , &
x_i &= \frac{m_i^2}{m_K^2} \ , &
U_i^2 &= \abs{c_{\ell i} - \frac{v c_{\nu i}}{m_i}}^2 \ .
\end{align}
When combined with the master formula \eqref{eq:K to ell master formula}, this result is consistent with the model-independent formula given in \cite[section 6 of][]{Arina:2021nqi}.

This computation exemplifies a second use of the factorization procedure:
In addition to facilitating model-independent constraints on the coupling to hidden sectors, it can also help streamline the derivation of more specific model-dependent constrains.
Once the master-formula associated with a given observable has been computed, it does not need to be adjusted anymore, and in order to adapt the result for a new model, it is sufficient to recompute the form-factors $F_\ell(x_q)$.
As a final remark, we note that while the form factors $F_\ell(x_q)$ depends on the observable in question, the hidden current correlation matrices in \cref{eq:correlation matrices} do not.
They can be provided once and for all, and re-used to compute a wide array of form factors associated with different observables, further reducing the need for re-computing ingredients in order to adapt a known result for a new model or observable.
 
\section{Conclusion and outlook}
\label{sec:conclusion}

In this work, we have shown that inclusive hidden particle production rates approximately factorize according to relation \eqref{eq:inclusive rate factorization},
and derived recipes for computing the reduced matrix elements $M_d$ and the hidden currents $J^d$ as series of Feynman diagrams.
We illustrated the factorization procedure by considering decays $K^+ \to \ell^+ X$ of charged koans into charged leptons and a number of hidden particles.
The resulting model-independent master formula \eqref{eq:model independent decay width} improves the model-independent formula given in \cite{Arina:2021nqi}
and parametrizes the impact of hidden sectors in terms of a single form factor $F_\ell(x_q)$ that can be constrained in a largely model-independent fashion.
To illustrate how to constrain the form factor in practice, we have re-interpreted the analysis of \cite{NA62:2020mcv} and derived the model-independent bounds \eqref{eq:A bound,eq:B bound}.

If the factorization approach is combined with an appropriately general list of portal operators, which can be constructed in a systematic and consistent fashion using \eg the \PET framework \cite{Arina:2021nqi},
it correctly accounts for both light \emph{and} heavy new particles.
In this case, the factorization approach is strictly more general than the \EFT approach,
and we have argued that it provides a powerful tool for the model-independent interpretation of hidden sector searches.
\\

Using the present work as a foundation, there are many potentially interesting avenues for future investigation.

Since the factorization approach relies on lists of portal operators being provided as a necessary input,
it will profit greatly from further efforts of extending the \EFT approach to also account for light new particles.
In particular, the \PET framework is well suited for providing the needed lists of portal operators, and constructing further \PETs will allow the factorization approach to be applied to a significantly larger variety of observables.
At this time, there are \PETs that extend the full \SM, \LEFT, which describes the physics of the light \SM fields \cite{Fermi:1934sk, Jenkins:2013zja, Jenkins:2013wua, Alonso:2013hga}, and \cPT, which describes the physics of the light pseudoscalar mesons \cite{Weinberg:1966fm, Weinberg:1968de, Cronin:1967jq, Schwinger:1967tc, Wess:1967jq, Dashen:1969ez, Gasiorowicz:1969kn}, by
coupling them to a single light new particle of spin 0, \textfrac12, or 1 \cite{Arina:2021nqi}.
These \PETs also account for the possibility of \SM fields coupling to multiple hidden fields with the \emph{same} spin, but they are not sufficient for describing a situation in which the \SM couples to multiple hidden particles with \emph{different} spins.
Hence, it would be useful to costruct a corresponding set of \PETs that include the relevant additional portal operators.

Likewise, it would be interesting to construct \PETs that extend \EFTs that capture different regimes of the \SM.
For instance, constructing \PETs that extend
\HQET, which captures the physics of heavy non-relativistic quarks \cite{Isgur:1989vq, Isgur:1989ed, Shifman:1987rj, Grinstein:1990mj, Georgi:1990um, Falk:1990yz},
and \SCET, which captures the physics of light but highly energetic particles \cite{Bauer:2000ew, Bauer:2000yr, Bauer:2001ct, Bauer:2003pi, Beneke:2004in, Bosch:2004th, Bauer:2008jx},
would make it possible to apply the factorization approach to hidden particle production in $B$ and $D$ meson decays,
where $B$ meson decays are of particular interest due to the persistent $B$ anomalies \cite{Descotes-Genon:2013wba,Altmannshofer:2013foa,LHCb:2017avl,LHCb:2021trn,Alda:2021ruz}.

On the hidden current side, it would be of great use to tabulate hidden current correlation matrices for a number of popular hidden sector models that include particles such as \ALPs, \HNLs, and dark photons.
Since the expressions for the hidden current correlation matrices are observable independent, these tabulated expressions can be used as input for a wide array of master formulae,
and could help facilitate \eg global parameter scans that combine constraints from a wide range of observables.
Finally, it would be interesting the study the renormalization scale dependence of the reduced matrix elements $M_d$ and the hidden currents $J^d$,
since a robust understanding of this running is necessary in order to combine constraints from observations at multiple characteristic energy scales.

\section*{Acknowledgments}

The author thanks Dr. Chiara Arina, Prof. Marco Drewes, and Dr. Jan Hajer for interesting discussions and providing valuable feedback regarding the contents of this paper.
The work was funded by the \SNF under grant \no{200020B-188712}.

\appendix
\appendixpage
\addappheadtotoc

\printbibliography

\end{document}